\documentclass[aps,a4paper,twocolumn]{revtex4-2}
\usepackage{amsmath}
\usepackage{graphicx}
\usepackage{epsfig}
\usepackage{bm}
\usepackage{xcolor}
\usepackage[colorlinks=true,
            linkcolor=blue,
            citecolor=blue,
            urlcolor=blue]{hyperref}
\usepackage{natbib}
\usepackage{caption}
\usepackage{subcaption}

\begin{document}
\title{Thermodiffusive coupled-transport phenomena in dense quark matter}
\author{Kamaljeet Singh}
\author{Kangkan Goswami}
\author{Raghunath Sahoo}
\email{Corresponding Author: Raghunath.Sahoo@cern.ch}
\affiliation{Department of Physics, Indian Institute of Technology Indore, Simrol, Indore 453552, India}
\date{\today}
\begin{abstract}

Coupled-transport phenomena reveal that heat, charge, and particle flows are intrinsically interconnected, providing deeper insight into the microscopic dynamics of a medium than independent transport processes. We study the behavior of the coupled-transport coefficients in hot and dense quark matter within the framework of the 2+1 flavor Nambu--Jona--Lasinio model at finite temperature and quark chemical potential. These coefficients characterize coupled-transport phenomena, where particle diffusion is driven by temperature gradients (Soret effect) and heat flow is induced by gradients in chemical potential (Dufour effect). These coefficients are estimated by solving the relativistic Boltzmann transport equation using the relaxation time approximation with temperature-dependent cross sections. We study the scaled Soret and Dufour coefficients as functions of temperature and quark chemical potential across the QCD phase diagram. We aim to understand the intricate behavior of the coupled-transport coefficients near the chiral symmetry restoration region. Our results indicate that coupled-transport coefficients are sensitive to the chiral phase transition and provide the first systematic insight into the cross-coupled-transport properties in dense quark matter.

\end{abstract}
\maketitle

\section{Introduction}

The primary objectives of the Large Hadron Collider (LHC) at CERN and the Relativistic Heavy Ion Collider (RHIC) at Brookhaven National Laboratory are to recreate the extreme conditions of temperature and energy density that prevailed in the early universe, enabling the investigation of strongly interacting matter under such conditions~\cite{Busza:2018rrf}. Under these extreme conditions, hadronic matter undergoes a transition to a deconfined state of quarks and gluons known as the quark-gluon plasma (QGP). Lattice Quantum Chromodynamics (lQCD) calculations suggest that at low baryon chemical potential, the transition from hadronic matter to QGP is a crossover, while at high baryon chemical potential and low temperature, a first-order phase transition is expected~\cite{Bellwied:2013cta, Bellwied:2017ttj, HotQCD:2018pds}. These two phases of strongly interacting matter are believed to be connected through a critical endpoint in the QCD phase diagram. In this context, investigating the transport and thermodynamic properties~\cite{Pradhan:2023rvf, Khaidukov:2019icg, Koothottil:2018akg, Goswami:2023eol, Sahoo:2023vkw, Gavin:1985ph, K:2022pzc} of strongly interacting matter across the phase transition becomes crucial for characterizing the evolution of the QGP medium.

Furthermore, the space-time evolution of the hot and dense matter created in heavy-ion collisions is strongly influenced by transport phenomena~\cite{Singh:2023pwf, Dey:2020awu, Ghosh:2019ubc, Nam:2012sg, Hattori:2016cnt, Hattori:2016lqx, Harutyunyan:2016rxm, Kerbikov:2014ofa, Feng:2017tsh, Wang:2020qpx, Tuchin:2012mf, Tiwari:2026vef, Dwibedi:2025boz}. Due to the rapid expansion of the fireball, the gradients~\cite{Singh:2024emy, Li:2019bgc, PhysRevB.105.235116, Abhishek:2020wjm, Singh:2025rwc, Singh:2025geq} in temperature, chemical potential, and flow velocity develop, leading to dissipative processes such as diffusion~\cite{Goswami:2023hdl, Goswami:2024hfg, Pradhan:2022gbm}, heat conduction~\cite{Singh:2023pwf, Singh:2023ues}, and viscosity~\cite{Huang:2011dc, Huang:2009ue, Agasian:2011st, Ghosh:2018cxb, Nam:2013fpa}. In particular, diffusion processes determine how conserved charge fluctuations propagate and survive until freeze-out, making transport coefficients important quantities in understanding the dynamics of strongly interacting matter~\cite{Shuryak:2000pd}. In addition, in systems with multiple thermodynamic forces, gradients of temperature and chemical potential can simultaneously generate particle and heat currents, leading to coupled-transport phenomena governed by the Onsager relations. Two important cross-coupled transport phenomena that arise in such systems are the Soret and Dufour effects. The Soret effect describes particle diffusion induced by a temperature gradient, while the Dufour effect describes heat flow induced by a chemical potential gradient. These coupled-transport coefficients provide important insight into the interplay between heat transport and particle diffusion in strongly interacting matter~\cite{Singh:2025dxh}. The Soret and Dufour effects have been extensively investigated in diverse areas of science and engineering due to their relevance in coupled heat and mass transport phenomena~\cite{rasool2020consequences}. In chemical and geophysical systems, these effects are known to significantly influence transport behavior, particularly under non-equilibrium conditions. For example, studies have demonstrated that the interplay between thermal and concentration gradients can strongly affect natural convection processes involving evaporation and condensation, where simultaneous heat and mass transfer occurs through species interdiffusion~\cite {jiang2020physical}. Furthermore, investigations employing the lattice Boltzmann approach with multiple-relaxation-time mechanisms have revealed that the combined action of Soret and Dufour mechanisms substantially modifies dual-diffusive convective flows~\cite{liu2019multiple}. Beyond classical fluids, analytical studies in magnetized plasma systems have emphasized the importance of these coupled-transport effects in electrodynamic environments~\cite{garcia2007dufour}. Extending such investigations to strongly interacting matter offers a promising direction for understanding non-equilibrium coupled-transport phenomena in QCD matter.

In this work, we study the behavior of the Soret and Dufour coefficients in strongly interacting dense quark matter using the 2+1 flavor Nambu--Jona--Lasinio (NJL) model at finite temperature and quark chemical potential. The NJL model is an effective model of Quantum Chromodynamics that describes spontaneous chiral symmetry breaking and dynamical mass generation of quarks. The model has been widely used to study the thermodynamic and transport properties of strongly interacting dense matter at finite temperature and chemical potential~\cite{Carlomagno:2025ayh, Zhang:2024apc, Ghosh:2025zkk, Dwibedi:2025boz}. In the NJL model, the constituent quark mass is generated dynamically due to chiral symmetry breaking. As temperature or chemical potential increases, the chiral condensate decreases, and chiral symmetry is partially restored, leading to a rapid change in the constituent quark masses near the chiral phase transition~\cite{Gastineau:2001zke, Shi:2015ufa}. Since transport coefficients depend on particle masses, thermodynamic quantities, and relaxation times, significant changes in transport properties are expected near the chiral phase transition. Here, using the constituent quark masses obtained from the NJL model, we estimate the coupled-transport coefficients using the relaxation time approximation with temperature-dependent cross sections. For the very first time, we investigate the behavior of the Soret and Dufour coefficients as functions of temperature and quark chemical potential across the chiral phase transition in dense quark matter following the NJL model. This helps establish coupled-transport coefficients as promising probes of the chiral phase transition by revealing their sensitivity to the underlying changes in the microscopic structure and transport dynamics of strongly interacting quark matter.

The paper is organized as follows: Sec.~\ref{Sec:Formalism} briefly discusses the 2+1 flavor NJL model and the thermodynamic formalism used in this work. This section includes the derivation of the Soret and Dufour coefficients by solving the Boltzmann transport equation (BTE) under the relaxation-time approximation (RTA). In Sec.~\ref{Sec:Results}, we discuss the behavior of the coupled-transport coefficients in the temperature--chemical potential plane and analyze their behavior near the chiral phase transition. Finally, Sec.~\ref{Sec:Summary} summarizes our work with possible outlooks.

\section{Formalism}\label{Sec:Formalism}
In this section, we discuss the Nambu--Jona--Lasinio model that we use to calculate the coupled-transport coefficients of hot and dense quark matter. Moreover, this section highlights the calculations for both Dufour and Soret coefficients in the framework of kinetic theory. Here, we solve the relativistic Boltzmann transport equation under the relaxation time approximation.

\subsection{Nambu--Jona--Lasinio model}
The Nambu--Jona--Lasinio model is an effective model of Quantum Chromodynamics that describes spontaneous chiral symmetry breaking and dynamical mass generation of quarks~\cite{Nambu:1961tp, Nambu:1961fr, Hatsuda:1994pi, Klevansky:1992qe}. In this work, we use the 2+1 flavor NJL model to estimate the constituent quark masses, which are then used to evaluate the coupled-transport coefficients.

The NJL Lagrangian for 2+1 flavors is given as~\cite{Klevansky:1992qe},
\begin{align}
\mathcal{L}_{NJL} &= \bar{q}(i\gamma^{\mu}\partial_{\mu} - \hat{m})q 
+ G \sum_{a=0}^{8} \left[ (\bar{q}\lambda^{a}q)^{2} + (\bar{q}i\gamma_{5}\lambda^{a}q)^{2} \right] \nonumber\\
&- K \left\{ \det[\bar{q}(1+\gamma_{5})q] + \det[\bar{q}(1-\gamma_{5})q] \right\},
\end{align}
where $q=(u,d,s)$ is the quark field, $\hat{m} = \text{diag}(m_u, m_d, m_s)$ is the current quark mass matrix, $\lambda^a$ are the Gell-Mann matrices in flavor space, and $G$ and $K$ represent the coupling constants for the four-point and six-point interactions, respectively. In this work, we use the parameter set from Ref.~\cite{Rehberg:1995kh}, where $\Lambda = 602.3$ MeV is a momentum cut-off scale, $G\Lambda^2 = 1.835$, $K\Lambda^5 = 12.36$, $m_u = m_d = 5.5$ MeV, and $m_s = 140.7$ MeV.

In the mean-field approximation, chiral symmetry breaking leads to the formation of quark condensates $\langle \bar{q}q \rangle$, which generate constituent quark masses dynamically. The gap equation for the constituent quark mass for each flavor can be written as~\cite{Rehberg:1995kh, Buballa:2003qv},
\begin{equation}
M_i = m_i - 4G \sigma_i + 2K \sigma_j \sigma_k,
\end{equation}
where $i,j,k = u,d,s$ in cyclic order, $M_i$ is the constituent quark mass, $m_i$ is the current quark mass, and $\sigma_i = \langle \bar{q}_i q_i \rangle$ is the quark condensate.
The quark condensate for each flavor at finite temperature and chemical potential is given by~\cite{Rehberg:1995kh, Buballa:2003qv},
\begin{equation}
\sigma_i = - 2N_c \int_{0}^{\Lambda} \frac{d^3\lvert\vec{k}_i\rvert}{(2\pi)^3}
\frac{M_i}{\omega_i}
\left[ 1 - f_i^0 - \bar{f}_i^0 \right],
\end{equation}
where $\omega_i = \sqrt{\vec{k}_i^2 + M_i^2}$ is the single particle energy, and $f_i^{0}(\bar{f_i^0})$ are the Fermi-Dirac distribution functions for quarks (antiquarks) given by
\begin{equation}\label{distribution}
f_i^{0}(\bar{f_i^0}) = \frac{1}{1 + \exp\left[\frac{\omega_i - \xi_i\mu_{q(\bar{q})}}{T}\right]},
\end{equation}
where, $\mu_q =\mu_B/3$ is the quark chemical potential and $\xi_{i} = \pm1$ for qurak and antiquark respectively. The constituent quark masses obtained from the gap equation are used to estimate the number density and coupled-transport coefficients in the medium. The relaxation time for quarks is estimated from the scattering rate and can be expressed as,
\begin{equation}
(\tau_R^{i})^{-1} = \sum_j n_j \, \sigma_{ij},
\end{equation}
where $n_j$ is the number density of quark flavor $j$ and $\sigma_{ij}$ is the scattering cross section between quark flavors $i$ and $j$. The scattering cross section is estimated using the relation between shear viscosity and cross section in a partonic medium. Using the theoretical lower bound $\eta/s = 1/(4\pi)$, the total isotropic cross section can be approximated as $\sigma_{\rm tot} \approx 0.716/T^2$~\cite{Fotakis:2019nbq, Xu:2007ns, Bouras:2009nn, Kovtun:2004de}. This temperature-dependent cross section is used to estimate the relaxation time for quarks in the medium.

\subsection{Coupled-transport coefficients}

In a medium with finite baryon density, the transport of baryon number and heat can be described within linear response theory. The particle diffusion current $\vec{j}$ and the heat current $\vec{I}$ can be written in terms of gradients of temperature and chemical potential as~\cite{Singh:2025dxh},
\begin{equation}~\label{matrix}
\begin{pmatrix}
\vec{j} \\
\vec{I}
\end{pmatrix}
=
- \begin{pmatrix}
D & S_T \\
D_F & \kappa
\end{pmatrix}
\begin{pmatrix}
\vec{\nabla} \mu_q \\
\vec{\nabla} T
\end{pmatrix},
\end{equation}
where $D$ is the diffusion coefficient, $S_T$ is the Soret coefficient, $D_F$ is the Dufour coefficient, and $\kappa$ is the thermal conductivity.

To evaluate the coupled-transport coefficients, we use the kinetic theory approach within the relaxation time approximation. The total distribution function for species $i$ is written as
\begin{equation}
f_i = f_i^0 + \delta f_i,
\end{equation}
where $f_i^0$ is the equilibrium distribution function given in Eq.~\eqref{distribution} and \(\delta f_i\) corresponds to the nonequilibrium correction generated by thermodynamic gradients.

To investigate nonequilibrium transport processes, we employ the relativistic Boltzmann transport equation within the relaxation time approximation. In the local rest frame and neglecting external forces, the kinetic equation for particle species \(i\) takes the form~\cite{Singh:2023pwf, Singh:2025dxh}
\begin{equation}\label{BTE-new}
\vec{v}_i \cdot \vec{\nabla} f_i 
= -\frac{\delta f_i(\vec{x}_i, \vec{k}_i)}{\tau_R^i}~,
\end{equation}
where \(\tau_R^i\) represents the relaxation time associated with species \(i\), while the particle velocity is given by \(\vec{v}_i=\vec{k}_i/\omega_i\). 

The spatial derivative of the equilibrium distribution function can be rewritten as
\begin{equation}\label{f0-grad-Gibbs-new}
\vec{\nabla} f_i^0
=
-\frac{\partial f_i^0}{\partial \omega_i}
\left[
\frac{\omega_i- \xi_i \mu_q}{T}\,\vec{\nabla}T
+
\xi_i\,\vec{\nabla}\mu_q
\right]~,
\end{equation}
which explicitly identifies the temperature and baryon chemical potential gradients as the driving forces that perturb the system away from equilibrium.

To determine the first-order correction to the distribution function, we introduce the following ansatz~\cite{Singh:2023pwf, Singh:2025dxh}
\begin{equation}\label{delta-f-new}
\delta f_i
=
(\vec{k}_i\cdot\vec{\Omega})
\frac{\partial f_i^0}{\partial \omega_i}~,
\end{equation}
where the vector \(\vec{\Omega}\) characterizes the medium response to thermodynamic gradients. Since the nonequilibrium transport is induced by gradients in temperature and baryon chemical potential, \(\vec{\Omega}\) can be expressed as
\begin{equation}\label{Omega-new}
\vec{\Omega}
=
\alpha\,\vec{\nabla}T
+
\beta\,\vec{\nabla}\mu_q~,
\end{equation}
with \(\alpha\) and \(\beta\) denoting the response coefficients associated with temperature and chemical potential gradients, respectively.

Substituting Eq.~\eqref{Omega-new} into Eq.~\eqref{delta-f-new}, we get
\begin{equation}\label{delta-f-final-new}
\delta f_i
=
\omega_i
\left[
\alpha(\vec{v}_i\cdot\vec{\nabla}T)
+
\beta(\vec{v}_i\cdot\vec{\nabla}\mu_q)
\right]
\frac{\partial f_i^0}{\partial \omega_i}~.
\end{equation}

By inserting Eq.~\eqref{delta-f-final-new} into Eq.~\eqref{BTE-new}, the Boltzmann equation can be rewrite as
\begin{align}
\frac{\omega_i- \xi_i \mu_q}{T}
(\vec{v}_i\cdot\vec{\nabla}T)
+
\xi_i
(\vec{v}_i\cdot\vec{\nabla}\mu_{q})
=
\frac{\omega_i}{\tau_R^i}
\Big[
\alpha
(\vec{v}_i\cdot\vec{\nabla}T)
\nonumber\\
+
\beta
(\vec{v}_i\cdot\vec{\nabla}\mu_{q})
\Big]~.
\end{align}

Equating the coefficients of independent thermodynamic gradients yields
\begin{align}
\alpha
=
\frac{\tau_R^i(\omega_i- \xi_i \mu_q)}
{T\omega_i},
\qquad
\beta
=
\frac{\xi_i\tau_R^i}{\omega_i}~.
\end{align}

Consequently, the nonequilibrium correction to the distribution function can be written as
\begin{equation}\label{delta}
\delta f_i = \frac{\vec{k}_i}{\omega_i}\cdot\tau_R^i
\left[ \frac{(\omega_i - \xi_i  \mu_q)}{T}\vec{\nabla}T + \xi_i \vec{\nabla}\mu_q \right]
\frac{\partial f_i^0}{\partial \omega_i}.
\end{equation}

According to kinetic theory, the particle diffusion current and the heat current can be written as~\cite{Singh:2025dxh}.

\begin{align}
\vec{j} &= \sum_i g_i \int \frac{d^3 \lvert \vec{k_i} \rvert}{(2\pi)^3 \omega_i} \vec{k}_i \, \delta f_i, \\
\vec{I} &= \sum_i g_i \int \frac{d^3 \lvert \vec{k_i} \rvert}{(2\pi)^3 \omega_i} (\omega_i - \xi_i  \mu_q)\vec{k}_i \, \delta f_i.
\end{align}

After substituting $\delta f_i$ and comparing with the transport matrix, given in Eq.~\eqref{matrix}, the transport coefficients can be written as
\begin{align}
S_T &= \sum_i g_i \frac{1}{3T^2} \int \frac{d^3 \lvert \vec{k_i} \rvert}{(2\pi)^3}
\frac{k_i^2}{\omega_i^2} (\omega_i - \xi_i  \mu_q)\tau_R^i f_i^0 (1 \pm f_i^0), \\
D_F &= \sum_i g_i \frac{1}{3T} \int \frac{d^3 \lvert \vec{k_i} \rvert}{(2\pi)^3}
\frac{k_i^2}{\omega_i^2} (\omega_i - \xi_i  \mu_q)\xi_i \tau_R^i f_i^0 (1 \pm f_i^0).\label{Eq:Dufour}
\end{align}

These expressions are used to estimate the coupled-transport coefficients in strongly interacting dense matter using the constituent quark masses obtained from the NJL model. In this article, we focus mainly on Soret and Dufour effects and their behavior near the chiral-phase transition. 

\section{Results}\label{Sec:Results}
\begin{figure*}
    \centering
    \includegraphics[width=0.45\linewidth]{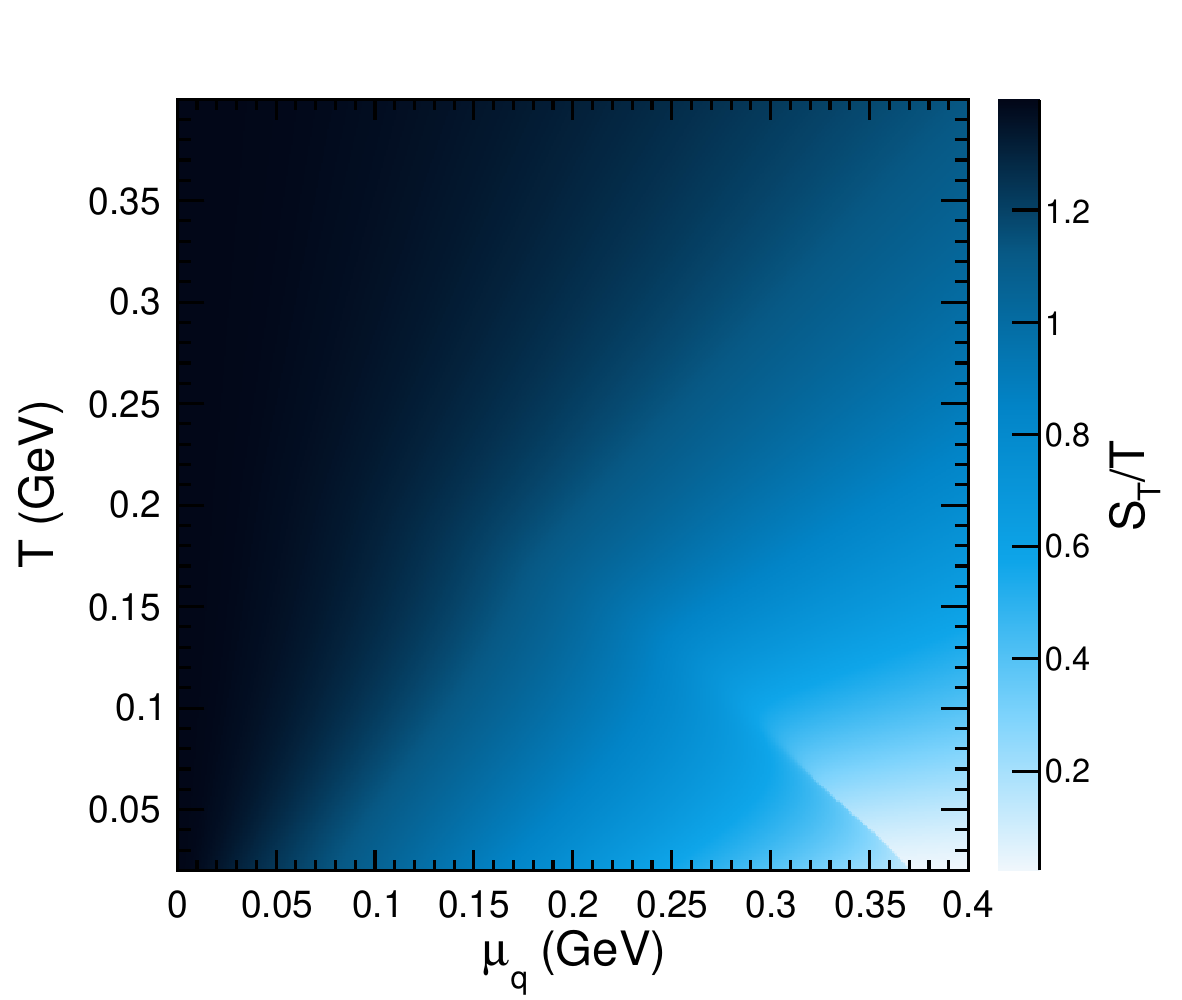}
    \includegraphics[width=0.45\linewidth]{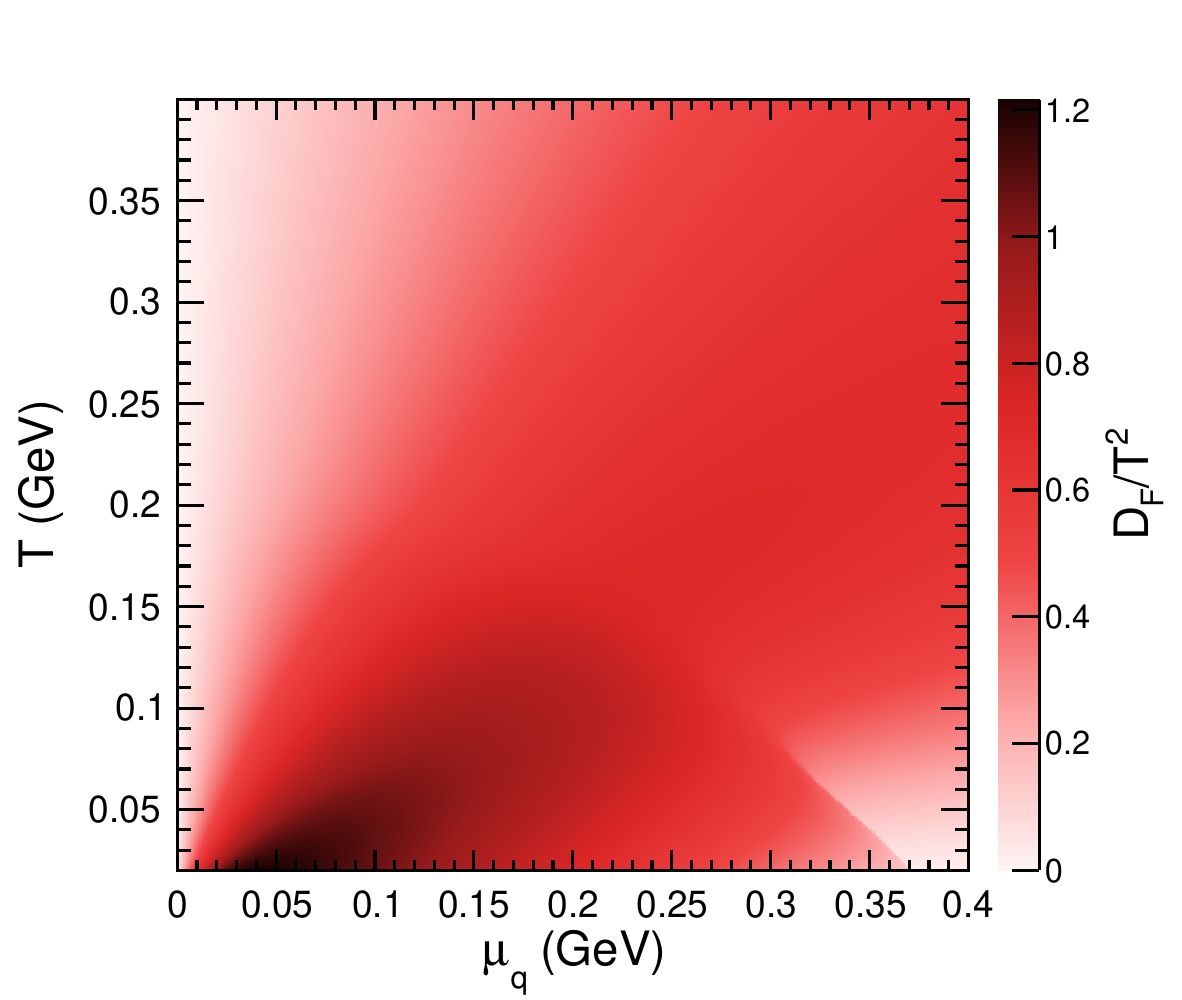}
    \caption{Contour plots of scaled Soret (left) and Dufour (right) coefficients in the $ (T-\mu_q)$ plane.}
    \label{Fig:S_Df_contour}
\end{figure*}
\begin{figure*}
    \centering
    \includegraphics[width=0.45\linewidth]{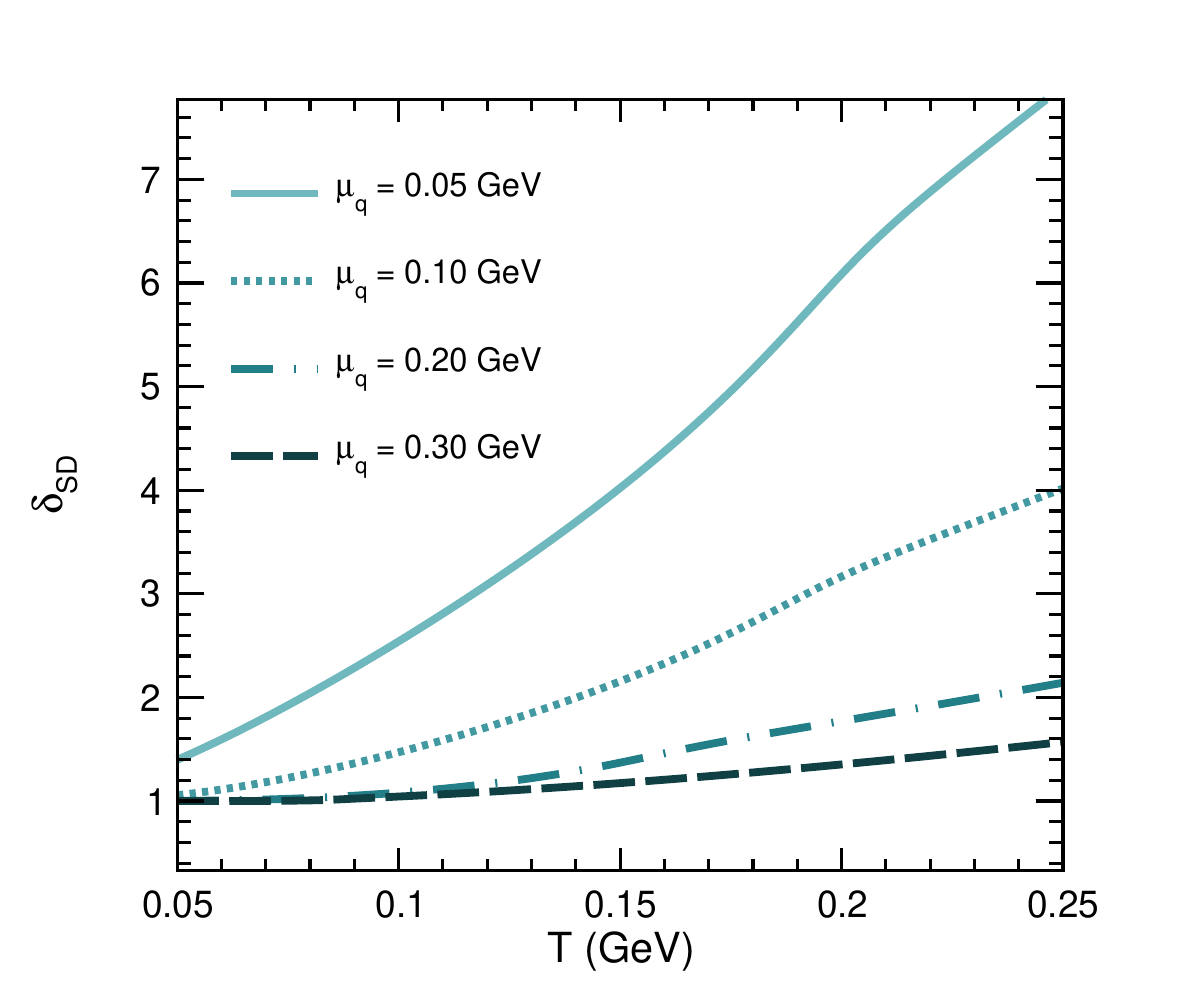}
    \includegraphics[width=0.45\linewidth]{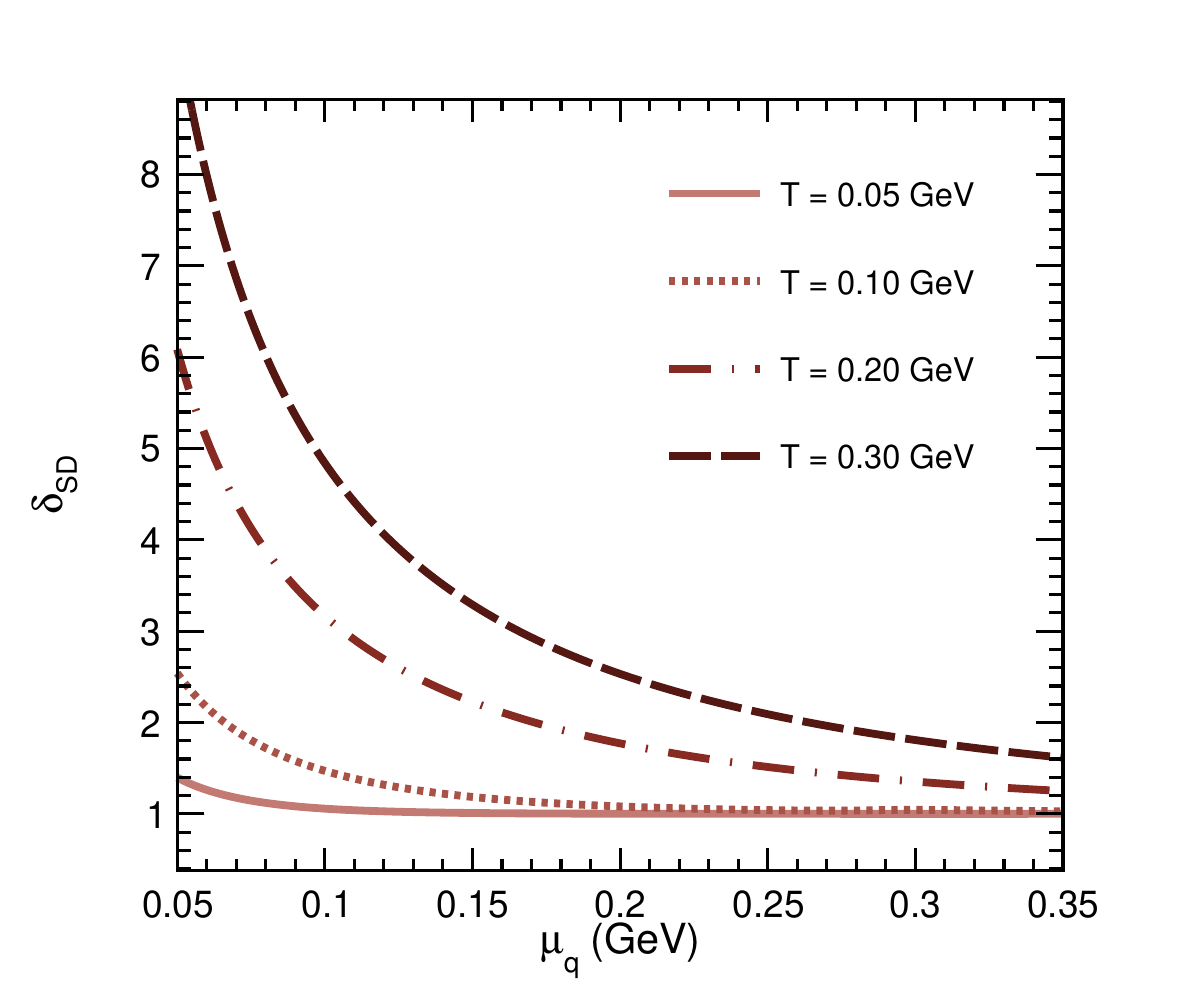}
    \caption{Ratio ($\delta_{SD} = S_{T}T/D_F$) of scaled Soret to scaled Dufour coefficient as a function of temperature (left) and quark chemical potential (right).}
    \label{Fig:Ratio_plot}
\end{figure*}

In this study, we investigate the coupled-transport coefficients of dense quark matter within the framework of the NJL model. Figure~\ref{Fig:S_Df_contour} presents the contour distributions of the scaled Soret coefficient $(S_{T}/T)$ and scaled Dufour coefficient $(D_{F}/T^{2})$ in the $T-\mu_{q}$ plane. Both coefficients exhibit a strong dependence on temperature and the quark chemical potential, reflecting the sensitivity of coupled-transport phenomena to the thermodynamic state of the medium and the underlying microscopic dynamics.
The left panel shows that the scaled Soret coefficient is comparatively larger in the low-density region and gradually decreases with increasing quark chemical potential. The temperature dependence is particularly weak at smaller $\mu_{q}$, whereas for the intermediate $\mu_{q}$ around (0.1 - 0.3) GeV, $(S_{T}/T)$ increases with temperature.  In contrast, the scaled Dufour coefficient displayed in the right panel exhibits a qualitatively different behavior. The coefficient remains strongly suppressed in the vanishing $\mu_{q}$ regime and increases with increasing quark chemical potential, reaching large values in the intermediate-temperature and finite-density region. This behavior arises due to the exact cancellation between quark and antiquark contributions at vanishing quark chemical potential, as governed by the $\xi_{i}$ factor appearing in Eq.~\eqref{Eq:Dufour}. We observe a strong enhancement of $D_F/T^2$ in the low-$T$ and intermediate-$\mu_{q}$ region, $T \sim 0.03{-}0.05$ GeV and $\mu_q \sim 0.04{-}0.10$ GeV. This behavior can be qualitatively understood from the approximate limit, $M \gg T$, which arises from the chirally broken phase at low temperature. Hence, it leads to the dependence of the scaled Dufour coefficient on an approximate functional form, $ D_{F}/T^{2} \propto \tanh(\mu_q/T)-\mu_q$. The competition between the increasing $\tanh(\mu_q/T)$ contribution and the explicit $-\mu_{q}$ term naturally generates a maximum at intermediate quark chemical potentials, leading to the observed enhancement of the Dufour coefficient in this region.

Furthermore, a noticeable change in both of the contour plots is observed in the region $T \sim 0.03 - 0.08$ GeV and $u_{q} \sim 0.28 - 0.36$ GeV, where both coefficients exhibit a sharp variation in their contour patterns. This region coincides with the first-order chiral phase transition in the NJL model and reflects discontinuity in the effective masses of the quarks. Since both coefficients depend explicitly on constituent quark masses, number densities, and relaxation times, the discontinuity of these quantities near the transition region strongly alters the transport response of the system. The observed contour structures therefore suggest that the coupled-transport coefficients may serve as useful probes of the chiral transition and the non-equilibrium transport dynamics of dense quark matter.

Figure~\ref{Fig:Ratio_plot} illustrates the ratio $\delta_{SD}=S_{T}T/D_F$, which quantifies the relative strength of thermodiffusion (Soret effect) to diffusive heat transport (Dufour effect). Physically, this ratio characterizes how efficiently the medium responds to a temperature gradient relative to a quark chemical-potential gradient. Larger values of $\delta_{SD}$ indicate that particle transport induced by thermal gradients dominates over heat transport driven by quark chemical-potential gradients, whereas smaller values signify an enhanced contribution from the Dufour effect. A value of $\delta_{SD}\simeq 1$ represents a regime where the coupled-transport phenomena become nearly comparable in magnitude, implying an equal contribution from $S_{T}$ and $D_{F}$ to diffusive and heat current, respectively.

The left panel depicts the temperature dependence of $\delta_{SD}$ at different quark chemical potentials. The ratio exhibits a systematic increase with temperature for all values of $\mu_{q}$, with the enhancement being more pronounced at lower quark chemical potentials. This behavior indicates that, as temperature increases, the medium becomes progressively more responsive to thermal gradients ($\vec{\nabla}T$) than to chemical-potential gradients ($\vec{\nabla}\mu_{q}$). In other words, thermodiffusive transport becomes increasingly dominant over diffusive heat flow. Physically, increasing temperature enhances quark diffusion through thermal gradients, while the relative contribution of concentration-gradient-driven heat transport becomes weaker. The particularly large values of $\delta_{SD}$ at small $\mu_{q}$ arise due to the suppression of the Dufour coefficient, where quark and antiquark contributions partially compensate each other. On the other hand, at relatively larger quark chemical potentials, the ratio gradually approaches $\delta_{SD} \simeq 1$, indicating that the Soret and Dufour contributions become comparable and the coupled-transport channels tend toward a more balanced response.

To understand the behavior of $\delta_{SD}$, we plot it as a function of quark chemical potential for fixed temperature values in the right panel of Fig.~\ref{Fig:Ratio_plot}. A rapid decrease in the ratio is observed with increasing $\mu_{q}$, followed by a gradual saturation to unity at larger chemical potentials. This trend signifies that increasing baryon asymmetry enhances the effectiveness of the Dufour effect relative to thermodiffusion. At low $\mu_{q}$, the medium predominantly responds to thermal gradients, resulting in strong thermodiffusive transport. However, as $\mu_{q}$ increases, quark chemical potential gradients become increasingly important and induce stronger heat transport, thereby reducing the relative dominance of the Soret effect. One should notice that the ratio saturates at different $\mu_{q}$ for different temperatures. For $T=0.05$ GeV, the ratio saturates around $\mu_{q} = 0.1$ GeV; however, for a higher temperature of around $T = 0.30$ GeV, the ratio only begins to saturate near $\mu_{q} = 0.35$ GeV. Mathematically, the saturation can be understood from the structure of the transport coefficients themselves. Both \(S_T\) and \(D_F\) arise from closely related microscopic transport integrals involving the same momentum-dependent kernel, constituent quark masses, relaxation times, and equilibrium distribution functions. The primary distinction originates from the additional factor \(\xi_i\) appearing in the Dufour coefficient, which explicitly incorporates the quark--antiquark asymmetry of the medium. At sufficiently large quark chemical potential, where the matter becomes increasingly baryon asymmetric and quark contributions dominate over antiquarks, the relative difference between the Soret and Dufour transport phenomena becomes less pronounced. Consequently, the gradual approach of $\delta_{SD}$ toward unity indicates a tendency toward a balanced competition between thermodiffusion and concentration-induced heat transport in dense quark matter.

\begin{figure*}
    \centering
    \includegraphics[width=0.45\linewidth]{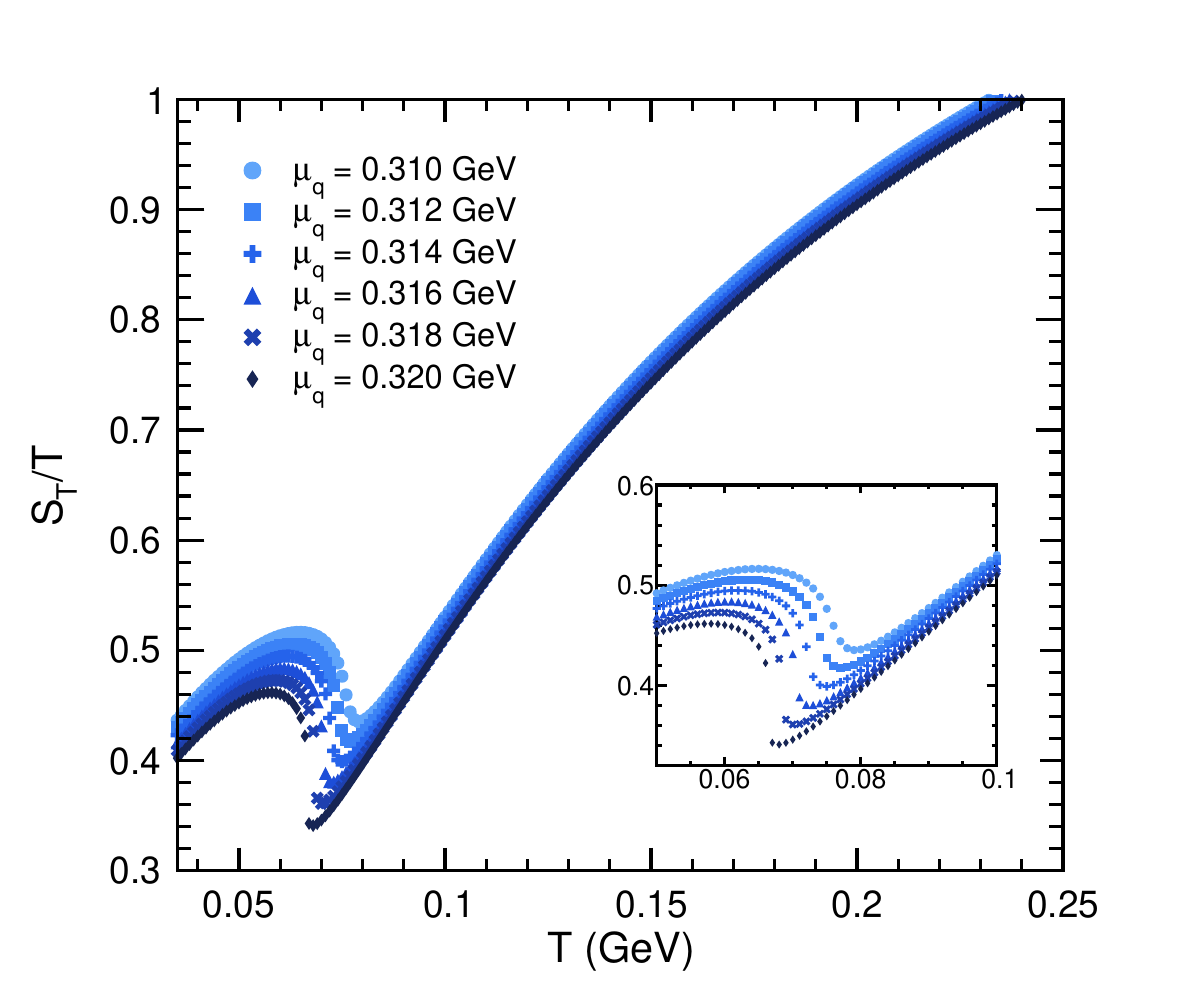}
    \includegraphics[width=0.45\linewidth]{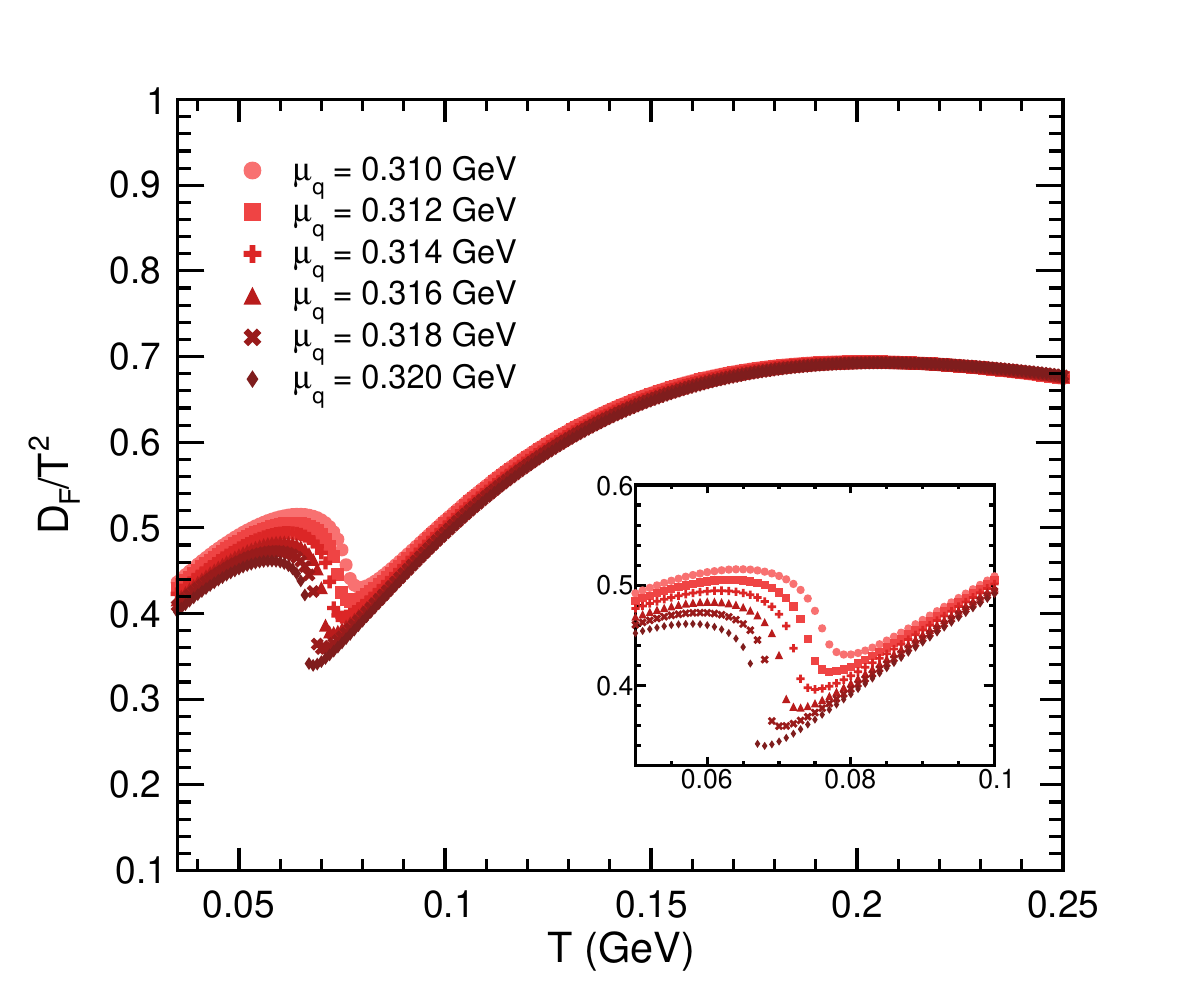}
    \caption{Temperature dependence of scaled Soret (left) and Dufour (right) coefficients at different values of quark chemical potential $\mu_{q}$.}
    \label{Fig:S_Df_T}
\end{figure*}

In Fig.~\ref{Fig:S_Df_T}, we present the temperature dependence of the scaled Soret coefficient and scaled Dufour coefficient for different values of quark chemical potential in the vicinity of the chiral phase transition region. Both coefficients exhibit a strong temperature dependence and become increasingly sensitive to the quark chemical potential near the chiral phase transition region. At relatively low temperatures ($T \lesssim 0.06$~GeV), both coefficients increase gradually with temperature. For example, the scaled Soret coefficient increases from approximately 0.42 to 0.50, while the scaled Dufour coefficient rises from about 0.40 to 0.50, depending on the value of $\mu_{q}$. As the temperature approaches the transition region, both coefficients exhibit a sharp dip within a narrow interval around $T \simeq 0.07-0.08$~GeV, after which they increase again with temperature. The magnitude of this dip increases with increasing $\mu_{q}$, in particular, at $\mu_q \simeq 0.32$~GeV, the scaled Soret coefficient dips from nearly 0.45 to about 0.35, while the scaled Dufour coefficient drops from approximately 0.44 to 0.34. Beyond the transition region, the scaled Soret coefficient rises continuously and reaches values close to 0.9 at $T \simeq 0.22$~GeV, whereas the scaled Dufour coefficient increases more gradually, attains a broad maximum around 0.69 near $T \simeq 0.20$~GeV, and subsequently exhibits a mild saturation at higher temperatures or a slight decrease at higher temperatures. Physically, this distinct behavior indicates that thermodiffusive transport remains increasingly efficient at high temperatures, whereas heat transport induced by carrier concentration gradients becomes comparatively less sensitive once the medium approaches a nearly chirally restored state.

The inset panels provide a magnified view of the transition region and clearly reveal that the dip-like structure becomes progressively sharper with increasing quark chemical potential and shows a discontinuity around $\mu_{q}\simeq 0.314$ GeV. Simultaneously, the dip systematically shifts toward lower temperatures, moving from approximately $T \simeq 0.078$~GeV at $\mu_q \simeq 0.31$~GeV to nearly $T \simeq 0.072$~GeV at $\mu_q \simeq 0.32$~GeV. A discontinuity begins to emerge around $\mu_q \simeq 0.314$~GeV, signaling increasingly rapid medium modifications near the transition boundary. This behavior reflects the movement of the chiral phase boundary in the $T-\mu_{q}$ plane within the NJL model framework and indicates a first-order transition region at higher baryon density. The observed sensitivity of both coefficients originates from their dependence on constituent quark masses, number densities, and relaxation times, all of which show discontinuous behavior near partial restoration of chiral symmetry. Furthermore, while the Soret coefficient remains finite for all values of $\mu_{q}$, the Dufour coefficient vanishes at $\mu_{q} = 0$ due to the exact cancellation between quark and antiquark contributions, implying that thermodiffusion remains active in baryon-symmetric matter, whereas concentration-gradient-driven heat transport becomes suppressed.

\section{Summary and Conclusion}\label{Sec:Summary}

In the present work, we carried out a systematic investigation of coupled-transport phenomena in dense quark matter by studying the Soret and Dufour effects within the framework of the 2+1 flavor Nambu--Jona--Lasinio model at finite temperature and quark chemical potential. Unlike conventional transport studies that focus on independent transport coefficients, the present analysis explores the reciprocal interplay between particle diffusion and heat transport in strongly interacting matter. Our results demonstrate that coupled-transport coefficients provide a unique direction for investigating the microscopic behavior of the medium near the chiral phase transition. In particular, the discontinuity of constituent quark masses associated with partial restoration of chiral symmetry strongly affects the transport response, producing characteristic transition behavior in both the Soret and Dufour coefficients. 

The key findings of the present study are summarized as follows:
\begin{itemize}
\item \textit{Strong sensitivity to the chiral phase transition}:  
Both the Soret and Dufour coefficients exhibit a discontinuity near the chiral transition region, which originates from the discontinuity in constituent quark masses, number densities, and relaxation time. The discontinuity and their systematic movement in the $T-\mu_{q}$ plane closely follow the evolution of the chiral phase boundary within the NJL model.

\item \textit{Distinct roles of thermodiffusion and diffusive heat transport}:  
The Soret coefficient remains finite throughout the explored phase space, demonstrating that thermodiffusion remains active even in baryon-symmetric matter. In contrast, the Dufour coefficient vanishes at $\mu_{q} = 0$ due to the exact cancellation between quark and antiquark contributions arising from the $\xi_{i}$ factor, implying that diffusive heat transport becomes suppressed in the absence of net baryon asymmetry.

\item \textit{Evidence of competing coupled-transport channels}:  
To quantify the competition between reciprocal coupled-transport, we introduced the ratio \begin{equation*}
    \delta_{SD}=\frac{S_{T}T}{D_{F}},
\end{equation*} which serves as a measure of the relative importance of thermodiffusion and diffusive heat transport. The analysis reveals that, in the low-density regime, the medium responds more efficiently to thermal gradients than to quark chemical-potential gradients, indicating dominant thermodiffusive transport. However, with increasing quark chemical potential, the Dufour response becomes progressively stronger, and $\delta_{SD}$ gradually approaches unity, signaling a more balanced competition between reciprocal coupled-transport channels.

\item \textit{Potential probe of QCD matter under extreme conditions}:  
The sensitivity of coupled-transport coefficients to temperature and quark chemical potential suggests that they may serve as novel probes of the chiral phase transition and nonequilibrium transport dynamics of strongly interacting matter. Such transport signatures could provide complementary insight into the evolution of hot and dense matter created in relativistic heavy-ion collisions and may also be relevant for dense astrophysical environments such as neutron stars.

\end{itemize}
 
The present study opens a promising direction toward understanding coupled transport in strongly interacting dense matter. A particularly interesting future extension would be the investigation of coupled-transport coefficients in strongly magnetized quark matter, where ultra-intense magnetic fields, such as those expected in magnetars, may induce anisotropic transport behavior and give rise to novel thermo-magnetic coupled-transport phenomena. Such studies could provide deeper insight into the interplay between heat, particle transport, and magnetic fields in QCD matter under extreme conditions.

\acknowledgments
K.S. acknowledges the doctoral fellowship from the UGC, Government of India. K.G. acknowledges the financial support from the Prime Minister's Research Fellowship (PMRF), Government of India. R.S. gratefully acknowledges the DAE-DST, Govt. of India funding under the mega-science project – “Indian participation in the ALICE experiment at CERN” bearing Project No. SR/MF/PS-02/2021-IITI (E-37123).

\end{document}